\documentclass[11pt,twoside]{article} 
\usepackage{cspm-asp2006}
\usepackage{epsfig,graphicx,natbib,url}  
\usepackage{lscape} 
\begin{document}
\setcounter{page}{115}

\markboth{Koza et al.}{Temporal Variations in Fibril Orientation} 
\title{Temporal Variations in Fibril Orientation} 
\author{J. Koza$^{1,2}$, 
        P. S\"{u}tterlin$^1$, 
        A. Ku\v{c}era$^2$ 
        and J. Ryb\'{a}k$^2$} 
\affil{$^1$Sterrekundig Instituut, Utrecht University, The Netherlands \\
       $^2$Astronomical Institute, SAS, Tatransk\'{a} Lomnica, Slovakia} 

\begin{abstract}  
  We measure variations in orientation of fourteen dynamic fibrils as
  a function of time in a small isolated plage and nearby network
  using a 10-min time sequence of H$\alpha$ filtergrams obtained by
  the Dutch Open Telescope.  We found motions with average angular
  velocities of the order of 1\,deg\,min$^{-1}$ suggesting systematic
  turning from one limit position to another, particularly apparent in
  the case of fibrils with lifetimes of a few minutes.  Shorter
  fibrils tend to turn faster than longer ones, which we interpret as
  due to vortex flows in the underlying granulation that twist
  magnetic fields.
\end{abstract}

\section{Introduction}
The solar chromosphere 
is filled with fibrils in plages and mottles in network seen on the
disk and with spicules seen at the limb.  Although the mutual
correspondence of fibrils, mottles, and spicules has not yet been
established directly, we believe that they represent the same feature
seen under different circumstances (cf.\
\cite{koza-Christopoulouetal2001}).  Their ubiquity is especially
evident in images taken at the center of strong spectral lines.  So
far most effort has been directed towards understanding their nature,
internal structure
\citep{koza-DePontieuetal2004,koza-Tziotziouetal2003,koza-Tziotziouetal2004},
and drivers \citep{koza-Hansteenetal2006,koza-DePontieuetal2007}. Less
attention has been paid to their less obvious tangential motions
(i.e., perpendicular to the axis) which may betray braiding of
chromospheric magnetic fields due to vortex granular flows underneath
\citep{koza-Brandtetal1988}.  The first spectroscopic observation of
spicule motions parallel to the limb was made by
\citet{koza-Pasachoffetal1968}.  \citet{koza-NikolskyandPlatova1971}
reported on the quasi-periodic motion of spicules along the limb with
tangential velocities of 10--15\,km\,s$^{-1}$ and amplitudes about
1\,arcsec.
\citet{koza-MamedovandOrudzhev1983a,koza-MamedovandOrudzhev1983b}
pointed out the similarity between radial (i.e., along the
line-of-sight) and tangential velocities of spicules along the limb
and speculated on the motion of spicules as a whole.

In this paper, we study tangential motions of fibrils in a small
isolated plage and nearby network using a 10-min time sequence of
H$\alpha$ filtergrams obtained with the Dutch Open Telescope (DOT).
We concentrate on relatively short-lived straight dynamic fibrils
(henceforth DFs, \cite{koza-deWijnandDePontieu2006}) exhibiting
conspicuous elongation and/or retraction within a few minutes.
Longer, more static or more curved fibrils are not considered here.
We for the first time present measurements of temporal variations in
DF orientations.  They suggest a relation between angular velocity and
DF length.

\begin{figure}
  \centering
  \includegraphics[width=0.7\textwidth]{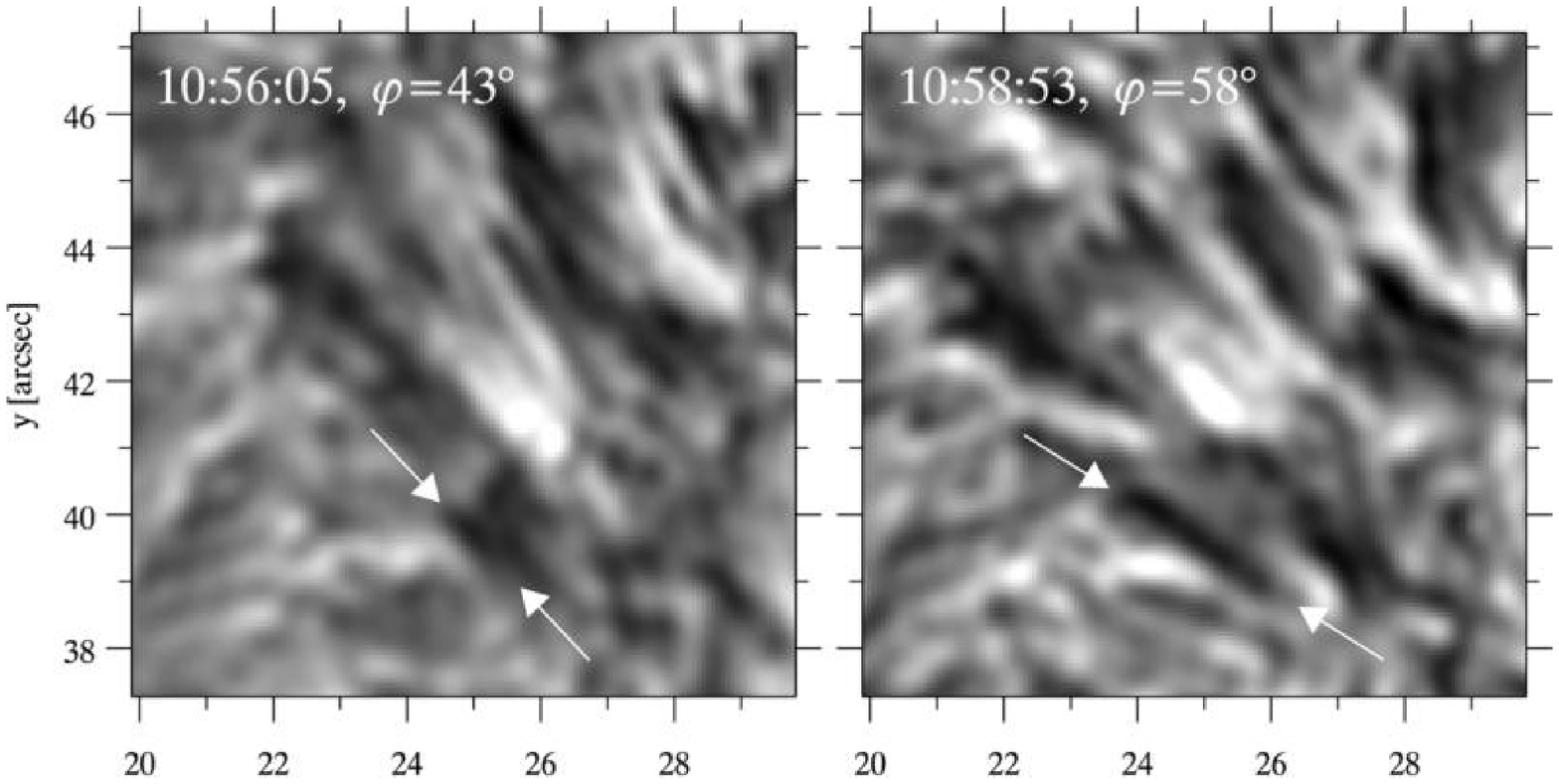}\\
  \vspace{-2mm}
  \includegraphics[width=0.7\textwidth]{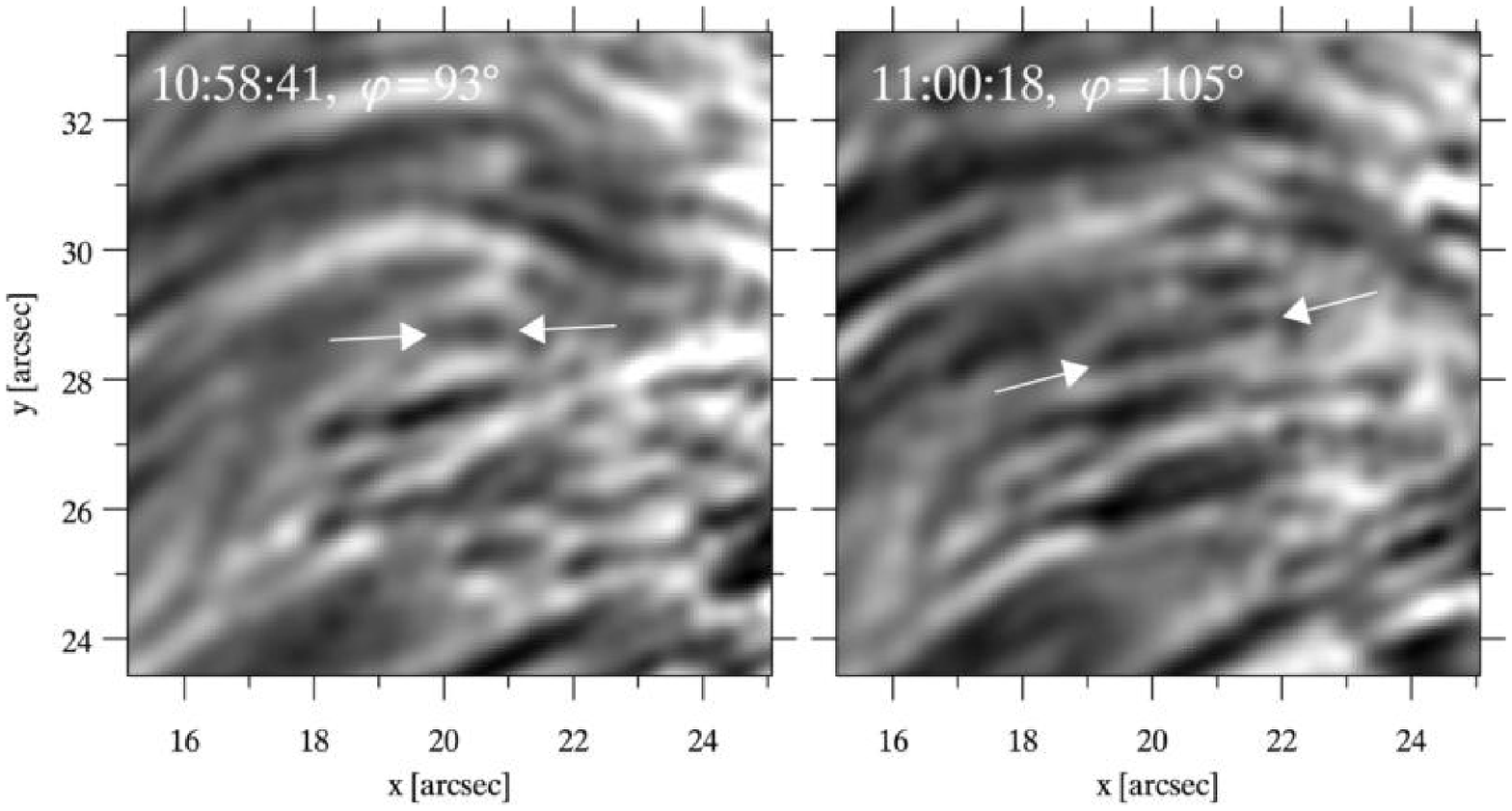}
  \caption[]{\label{koza-fig:orientations}
The orientations and lengths of two fibrils highlighted by arrows at
intervals of 2.8\,min (upper panels) and 1.6\,min (lower panels).  The
angle $\varphi$ is measured clockwise from the fibril to the righthand
$y$-axis.
}
\end{figure}

\section{Observations and Measurements}
We use data from the DOT obtained on April 24, 2006 for an isolated
plage at $\mu=0.768$.  A tomographic multiwavelength image sequence
was recorded during excellent seeing from 10:53:05~UT until
11:02:54~UT.  The seeing quality measured at the G band by the average, maximum, and
minimum Fried parameter $r_{\rm 0}$ was 13.0, 16.2, 8.6, respectively.  The
processed data and movies are available at {\tt
http://dotdb.phys.uu.nl/DOT/}.  The burst cadence was 12\,s.  The
resulting time sequence was reconstructed by speckle masking and other
steps as summarized in \citet{koza-Ruttenetal04}.  In this study, we
analyse the 10-min sequence of H$\alpha$ filtergrams taken by the
DOT Lyot filter \citep{koza-Gaizauskas1976} with a FWHM passband of
0.025\,nm at $\Delta\lambda=-0.03$\,nm from line center.

Inspecting the H$\alpha$ movie frame-by-frame we focused on
relatively short-lived ($\approx$5\,min), straight DFs exhibiting
apparent elongation or/and retraction.  We identified fourteen such
DFs and measured the image coordinates of the apparent feet and tops
per individual frame.  Because the feet locations also vary with time
we used the time-averaged feet positions as references.  We measured
the DF orientations with respect to the $y$-axis (terrestrial north)
and the DF lengths along lines connecting their tops and 
reference foot locations.
Figure~\ref{koza-fig:orientations} shows two examples.

\begin{figure}
  \centering \includegraphics[width=0.45\textwidth]{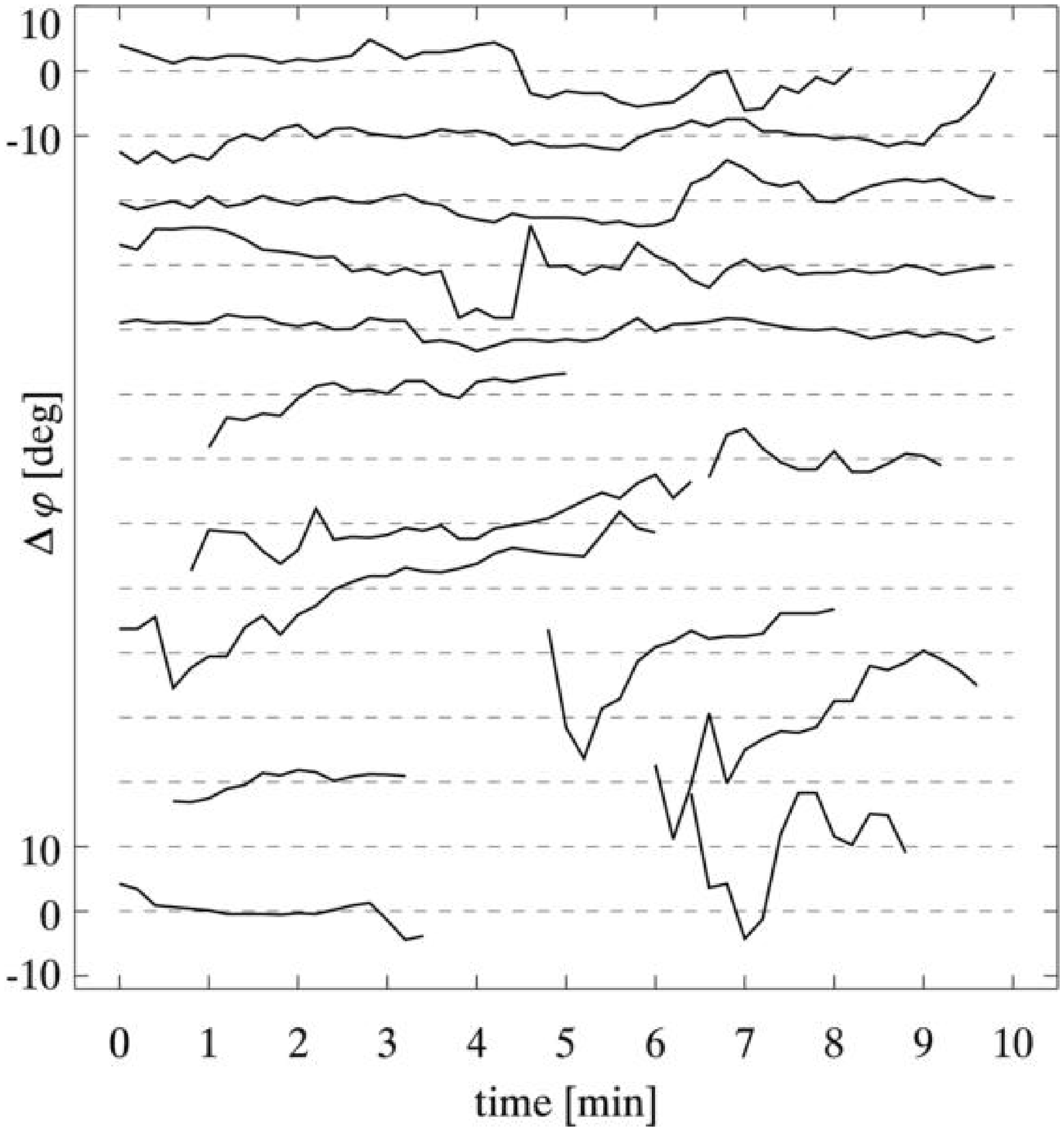}
  \includegraphics[width=0.45\textwidth]{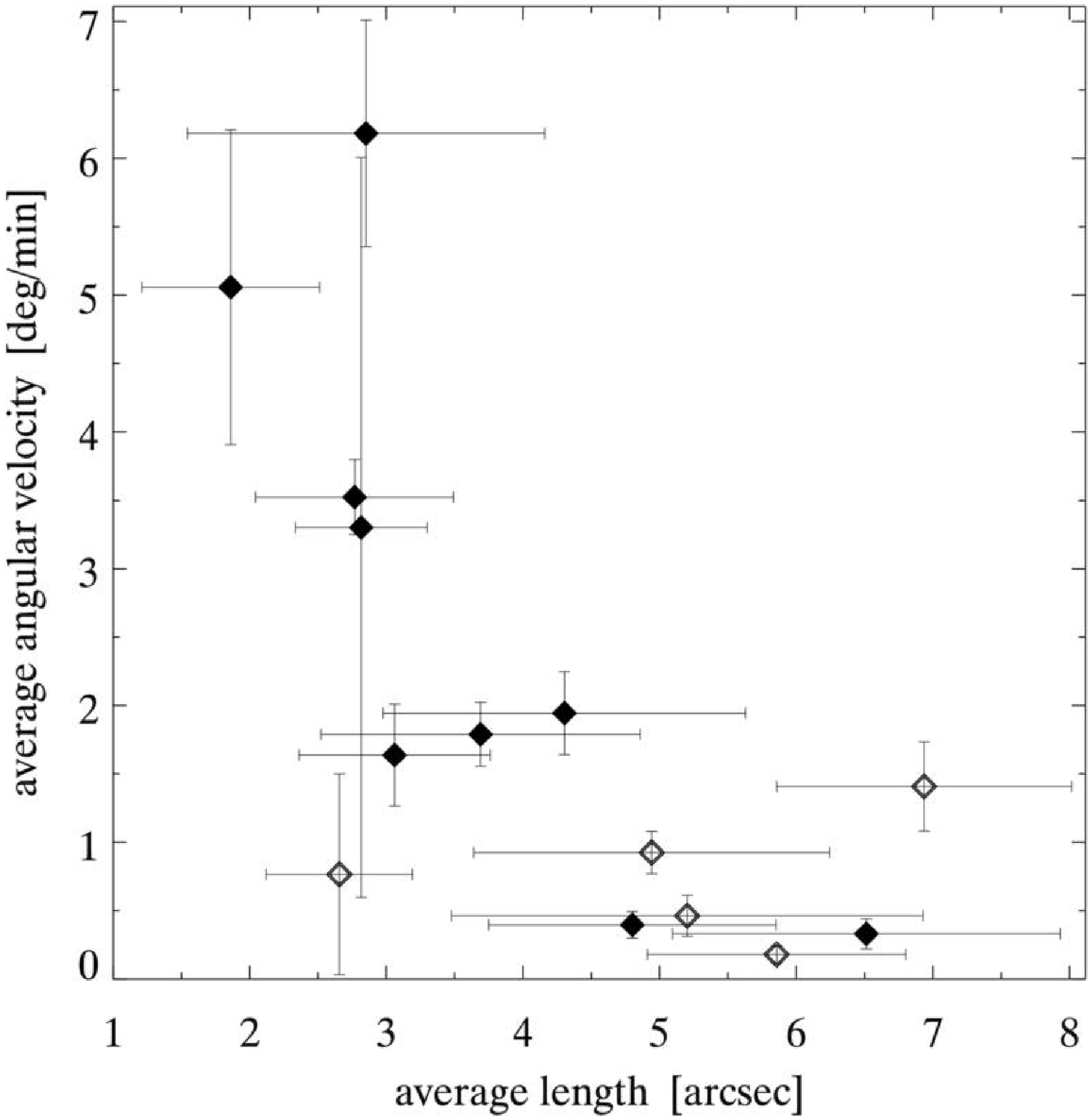}
  \caption[]{\label{koza-fig:dphi-omegal}
{\em Left:} Temporal variations in fibril orientation $\Delta \varphi$
 measured as difference with the average value represented by dashed
 lines.  The shifts of the curves in time correspond to the fibril
 occurrence within the image sequence.  {\em Right:} Average angular
 velocities of the orientation variation versus average fibril length.
 The filled and empty diamonds represent fibrils turning
 counterclockwise and clockwise, respectively.
}
\end{figure}

\section{Results}
Figure~\ref{koza-fig:dphi-omegal} shows the measured variations in
orientation for the fourteen selected DFs and the rate of change in
orientation inferred from linear fits to the curves at left plotted
against average DF length.  Although the measurements suffer from
uncertainty and subjectivity, systematic trends appear.  The
orientation variations of some short-lived DFs can be characterised as
a progression between sign changes, i.e., orientation change from one
limit position to another.  In contrast, the DFs present during the
whole time sequence show stable orientation with only episodic
deviations from the average value.  Figure~\ref{koza-fig:dphi-omegal}
suggests that the angular turning speed may be related to DF length,
and that shorter DFs turn faster than longer ones.  The average fibril
lengths and angular velocities yield an estimate of centrifugal
acceleration of 1\,m\,s$^{-2}$.

\section{Discussion} 
Because of projection and the magnetic nature of DFs
\citep{koza-DePontieuetal2004,koza-DePontieuetal2007,koza-Hansteenetal2006},
we interpret the temporal variation of the measured fibril orientation
as a sum of variations in azimuth and inclination of magnetic flux
tubes with respect to the local vertical.  In the context of
force-free fields \citep[e.g.][]{koza-LustandSchluter1954}, the
phenomenon may indicate field twisting and braiding by vortex granular
flows beneath \citep{koza-Brandtetal1988}, injection of twist into the
corona, and twisted coronal fans seen in TRACE EUV movies.  On the
basis of the indicated relation between angular velocity and length
(Fig.\,\ref{koza-fig:dphi-omegal}) we suggest that granulation with
larger vortical flows produce more upright flux tubes, i.e., faster
turning and shorter in projection on the disk, but better seen as long
spicules at the limb with conspicuous tangential motions as reported
in \citet{koza-Pasachoffetal1968},
\citet{koza-NikolskyandPlatova1971}, and
\citet{koza-MamedovandOrudzhev1983a,koza-MamedovandOrudzhev1983b}.  In
contrast, granulation with smaller or no vortical flows permits more
slanted flux tubes with smaller angular velocities, longer in on-disk
projection but hardly observable on the limb because of crowding.
Measurements of the horizontal flow fields in the photosphere and
elimination of the projection effects are needed to test this
scenario.  The observed angular velocity of $\approx
1$\,deg\,min$^{-1}$ implies more intensive twisting of the field lines
than in sunspots which have typical rotation velocities of about
1\,deg\,h$^{-1}$ \citep{koza-Kucera1982, koza-Brownetal2003}.

\section{Summary}
Using a time sequence of high-resolution H$\alpha$ filtergrams
obtained by the DOT we have searched for temporal variation in the
azimuthal orientation of fourteen dynamic fibrils.  They show
significant variation, for shorter-lived fibrils indicating turning
motions at about 1\,deg\,min$^{-1}$.  Shorter DFs turn faster than
longer ones, which may indicate difference in granulation vorticity.
This conjecture suggests measurements of horizontal flow fields in the
photosphere together with elimination of the projection effects in
fibril imaging.\\

\acknowledgements We thank Rob Rutten for improvements to the text.
  J.\,Koza's research is supported by an EC Marie Curie Intra European
  Fellowship. This research was part of the European
  Solar Magnetism Network and is supported by Slovak agency
  VEGA (2/6195/26).



\begin{thebibliography}{}

\bibitem[\protect\astroncite{{Brandt} et~al.}{1988}]{koza-Brandtetal1988}
{Brandt} P.~N., {Scharmer} G.~B., {Ferguson} S., {Shine} R.~A., {Tarbell}
  T.~D., 1988,
  Nature 335, 238

\bibitem[\protect\astroncite{{Brown} et~al.}{2003}]{koza-Brownetal2003}
{Brown} D.~S., {Nightingale} R.~W., {Alexander} D., {Schrijver} C.~J.,
  {Metcalf} T.~R., {Shine} R.~A., {Title} A.~M., {Wolfson} C.~J., 2003,
  \solphys  216, 79

\bibitem[\protect\astroncite{{Christopoulou}
  et~al.}{2001}]{koza-Christopoulouetal2001}
{Christopoulou} E.~B., {Georgakilas} A.~A., {Koutchmy} S., 2001,
  \solphys  199, 61

\bibitem[\protect\astroncite{{De Pontieu} et~al.}{2004}]{koza-DePontieuetal2004}
{De Pontieu} B., {Erd{\'e}lyi} R., {James} S.~P., 2004,
  Nature 430, 536

\bibitem[\protect\astroncite{{De Pontieu} et~al.}{2007}]{koza-DePontieuetal2007}
{De Pontieu} B., {Hansteen} V.~H., {Rouppe van der Voort} L., {van Noort} M.,
  {Carlsson} M., 2007,
  \apj~  655, 624

\bibitem[\protect\astroncite{{de Wijn} \& {de
  Pontieu}}{2006}]{koza-deWijnandDePontieu2006}
{de Wijn} A.~G., {de Pontieu} B., 2006,
  \aap~  460, 309

\bibitem[\protect\astroncite{{Gaizauskas}}{1976}]{koza-Gaizauskas1976}
{Gaizauskas} V., 1976,
  \jrasc\  70, 1

\bibitem[\protect\astroncite{{Hansteen} et~al.}{2006}]{koza-Hansteenetal2006}
{Hansteen} V.~H., {De Pontieu} B., {Rouppe van der Voort} L., {van Noort} M.,
  {Carlsson} M., 2006,
  \apjl\  647, L73

\bibitem[\protect\astroncite{{Kucera}}{1982}]{koza-Kucera1982}
{Kucera} A., 1982,
  Bulletin of the Astronomical Institutes of Czechoslovakia  33, 345

\bibitem[\protect\astroncite{{L{\"u}st} \&
  {Schl{\"u}ter}}{1954}]{koza-LustandSchluter1954}
{L{\"u}st} R., {Schl{\"u}ter} A., 1954,
  Zeitschrift f.\ Astrophys.\  34, 263

\bibitem[\protect\astroncite{{Mamedov} \&
  {Orudzhev}}{1983a}]{koza-MamedovandOrudzhev1983a}
{Mamedov} S.~G., {Orudzhev} E.~S., 1983a,
  Soviet Astronomy  27, 692

\bibitem[\protect\astroncite{{Mamedov} \&
  {Orudzhev}}{1983b}]{koza-MamedovandOrudzhev1983b}
{Mamedov} S.~G., {Orudzhev} E.~S., 1983b,
  \azh~  60, 1192

\bibitem[\protect\astroncite{{Nikolsky} \&
  {Platova}}{1971}]{koza-NikolskyandPlatova1971}
{Nikolsky} G.~M., {Platova} A.~G., 1971,
  \solphys  18, 403

\bibitem[\protect\astroncite{{Pasachoff} et~al.}{1968}]{koza-Pasachoffetal1968}
{Pasachoff} J.~M., {Noyes} R.~W., {Beckers} J.~M., 1968,
  \solphys  5, 131

\bibitem[\protect\astroncite{{Rutten} et~al.}{2004}]{koza-Ruttenetal04}
{Rutten} R.~J., {Hammerschlag} R.~H., {Bettonvil} F.~C.~M., {S{\"u}tterlin} P.,
  {de Wijn} A.~G., 2004,
  \aap\  413, 1183

\bibitem[\protect\astroncite{{Tziotziou} et~al.}{2003}]{koza-Tziotziouetal2003}
{Tziotziou} K., {Tsiropoula} G., {Mein} P., 2003,
  \aap\  402, 361

\bibitem[\protect\astroncite{{Tziotziou} et~al.}{2004}]{koza-Tziotziouetal2004}
{Tziotziou} K., {Tsiropoula} G., {Mein} P., 2004,
  \aap\  423, 1133

\end{thebibliography}
\end{document}